\documentclass[sigconf]{acmart}
\AtBeginDocument{%
  }

\copyrightyear{2024} 
\acmYear{2024} 
\setcopyright{rightsretained} 
\acmConference[CHI EA '24]{Extended Abstracts of the CHI Conference on Human Factors in Computing Systems}{May 11--16, 2024}{Honolulu, HI, USA}
\acmBooktitle{Extended Abstracts of the CHI Conference on Human Factors in Computing Systems (CHI EA '24), May 11--16, 2024, Honolulu, HI, USA}
\acmDOI{10.1145/3613905.3643983}
\acmISBN{979-8-4007-0331-7/24/05}

\begin{document}

\title{Human-Centered Privacy Research in the Age of Large Language Models}



\author{Tianshi Li}
\email{tia.li@northeastern.edu}
\orcid{0000-0003-0877-5727}
\affiliation{%
    \institution{Northeastern University}
    \city{Boston}
    \state{MA}
    \country{USA}
}

\author{Sauvik Das}
\email{sauvik@cmu.edu}
\affiliation{%
  \institution{Carnegie Mellon University}
  \city{Pittsburgh}
  \state{PA}
  \country{USA}
}

\author{Hao-Ping (Hank) Lee}
\email{haopingl@cs.cmu.edu}
\affiliation{%
  \institution{Carnegie Mellon University}
  \city{Pittsburgh}
  \state{PA}
  \country{USA}
}

\author{Dakuo Wang}
\email{d.wang@neu.edu}
\affiliation{%
    \institution{Northeastern University}
    \city{Boston}
    \state{MA}
    \country{USA}
}

\author{Bingsheng Yao}
\email{arthuryao33@gmail.com}
\affiliation{%
  \institution{Rensselaer Polytechnic Institute}
  \city{Troy}
  \state{NY}
  \country{USA}
}

\author{Zhiping Zhang}
\email{zhip.zhang@northeastern.edu}
\affiliation{%
  \institution{Northeastern University}
  \city{Boston}
  \state{MA}
  \country{USA}
}

\renewcommand{\shortauthors}{Li et al.}

\begin{abstract}
The emergence of large language models (LLMs), and their increased use in user-facing systems, has led to substantial privacy concerns.
To date, research on these privacy concerns has been model-centered: exploring how LLMs lead to privacy risks like memorization, or can be used to infer personal characteristics about people from their content.
We argue that there is a need for more research focusing on the human aspect of these privacy issues: e.g., research on how design paradigms for LLMs affect users' disclosure behaviors, users' mental models and preferences for privacy controls, and the design of tools, systems, and artifacts that empower end-users to reclaim ownership over their personal data.
To build usable, efficient, and privacy-friendly systems powered by these models with imperfect privacy properties, our goal is to initiate discussions to outline an agenda for conducting human-centered research on privacy issues in LLM-powered systems.
This Special Interest Group (SIG) aims to bring together researchers with backgrounds in usable security and privacy, human-AI collaboration, NLP, or any other related domains to share their perspectives and experiences on this problem, to help our community establish a collective understanding of the challenges, research opportunities, research methods, and strategies to collaborate with researchers outside of HCI.
\end{abstract}




\maketitle

\section{Background}

Large language models (LLMs) are transforming people's lives in many ways, but also present numerous risks --- and chief among these risks is privacy.
The NLP and system security communities have initiated extensive research into these models, focusing on the new privacy challenges they present and their capabilities for preserving user privacy.
One major problem is that these models can memorize and output training data~\cite{carlini2021extracting, carlini2022quantifying, zhang2021counterfactual}.
As the models are trained on vast amounts of data, including user data, this has raised new data leak risks.
For instance, research has found that prompting the model to continuously output ``poem'' can trick it into leaking training data verbatim~\cite{nasr2023scalable}.
Beyond memorization, LLMs can be used to extract personal attributes of individuals from seemingly harmless text~\cite{staab2023beyond}.
For example, given the text \textit{``I always get stuck there waiting for a hook turn''}, LLMs can help malicious actors infer that this person is in Melbourne because a hook turn is a traffic maneuver particularly used there. 
Research has also shown that LLMs lack the commonsense about social privacy norms, and have trouble keeping a secret~\cite{mireshghallah2023can} and that instruction-tuned models can be easily tricked by third-party adversaries to ignore privacy-protecting instructions \cite{chen2023can}.

Despite the privacy issues exhibited in these models and the lack of effective defensive methods, we are witnessing a rapidly growing trend of LLMs being integrated into interactive computing systems and placed in users' hands.
The most high-profile LLM application --- LLM-based conversational agents (CAs), such as ChatGPT --- are increasingly being incorporated into high-stakes application domains including healthcare~\cite{Leonard2023}, finance~\cite{Estrada2023, Ferreira2023, taver2023chatgpt}, and personal counseling~\cite{Germain2023, kimmel2023chatgpt}.
However, \citet{zhang2023s} found that the high utility of the tool and the human-like interactions encourage users to share sensitive and personally identifiable information with LLM-based CAs.
Despite this, users constantly face challenges in protecting their privacy due to the inherent tension between privacy and utility, their flawed mental models, and dark patterns in the design of privacy management features~\cite{zhang2023s}.
Given these challenges, we believe the HCI community has a responsibility to foster a paradigm shift in LLM-centered privacy research.
This shift should move from research that solely investigates the privacy risks entailed by a model, to research that empowers \textit{humans} to act on and specify their privacy preferences, when interacting with LLM-powered interactive systems, in a usable, convenient, efficient, and effective way.
Below are example questions that we hope our research community can explore:

\begin{itemize}
    \item What do users perceive as the privacy risks of LLMs? How do these perceptions align with the known risks?
    \item How can we promote data sovereignty in users' use of LLMs?
    \item How do users' (mis)perceptions of the privacy risks of LLMs affect their ability to manage their privacy when using applications/services powered by LLMs?
    \item How does the training of LLMs on vast web-scraped data affect users' privacy perceptions and behaviors in general (non-LLM) online disclosure?    
    \item How do different methods of designing interactions powered by LLMs affect users' awareness of how LLMs process their data and privacy-related behaviors?
    \item How can we educate the general public about the emerging privacy risks entailed by LLMs?
    
    \item How can we aid users in dealing with the trade-offs between privacy and other design deliberations (e.g., utility, convenience)?
    \item What are privacy challenges in personalized LLM-based agents, especially when they are integrated into socio-technical systems?
    
    \item How can we manage the tension between an individual user's privacy vs. other values of societal importance, such as safety and alignment efforts?
    \item How can regulatory efforts be designed to promote the development of privacy-respectful LLM-based systems?
    \item As more people start writing code with the help of LLMs, how do LLMs affect practitioners' abilities to handle privacy?
\end{itemize}

The primary goal of this Special Interest Group (SIG) is to bring together researchers with backgrounds in usable security and privacy, human-AI interaction, NLP, or any other related domains to collectively outline a research agenda to address the pressing and emergent privacy challenges entailed by large language models.
To facilitate concrete progress towards this goal, below we outline four key areas of focus for the SIG: Understanding Privacy Challenges for Users; Designing Privacy-friendly Interfaces of LLM-based Systems; Building Usable Tools for Privacy Management for LLM-based Systems; Challenges and Solutions Beyond Individual Users.

\section{Understanding Privacy Challenges for Users}
In LLM-based systems, one end-to-end model is usually expected to serve all the requests of varied use cases.
However, privacy concerns are contextualized and subjective, which means that studying privacy risks solely at the model level can yield an incomplete understanding of real-world issues.
Therefore, we believe it is important to investigate research questions such as: What is the impact, on user privacy, of interactions with LLMs in different contexts? What do users perceive as the most significant risks? And how do these perceived risks align or differ from our understanding of the actual risks?
This situated, human-centered research method can offer complementary insights to model-centered research for designing privacy-respectful LLM-based systems.
Taking research on the LLM memorization risks as an example, experiments with different models can reveal that memorization positively correlates with the size of the model and the frequency of text occurrence~\cite{carlini2022quantifying}.
Furthermore, user interviews have revealed that users are more aware and concerned about the risks of memorization when they use ChatGPT to revise original writings, such as novels and research papers, due to concerns about idea theft~\cite{zhang2023s}.
The former finding assists model developers in estimating general risks, while the latter is instrumental in designing interfaces that alert users to specific privacy risks when handling tasks with significant privacy implications.

LLMs are trained on tremendous amounts of web scraped data.
This suggests a inherently surveillant nature of LLMs, which means that the privacy impact is not limited to direct interactions with LLMs, but can also occur in other non-LLM-related online disclosures.
The lack of transparency of training data of the proprietary LLMs has become a focal problem in the ML community, while there is relatively less discussion on the user privacy aspect.
How does training LLMs on vast web-scraped data impact the risks of users' general online disclosure? How can we assist users in understanding and preventing these risks?
Moreover, as people frequently disclose their communications in private (e.g., emails) or semi-private (e.g., Facebook group posts) contexts with others to ChatGPT (discussed as the interdependent privacy issues in \citet{zhang2023s}), to what extent can this affect people's perceptions of privacy and the social dynamics of online activities?

\section{Designing Privacy-friendly Interfaces of LLM-based Systems}
In this aspect, we are interested in one main question: How do different types of interactions with LLMs affect users' privacy-related mental models and behaviors?
In addition to LLM-based conversational agents, there are other applications built with LLMs that afford other LLM-powered interactions.
For example, one less explicit type of LLM-powered interaction is embedding LLM-based autocompletion in a text editor in a web browser or other desktop applications such as GitHub Copilot.
Some applications employ less direct interactions between users and LLMs, including real-world products (e.g., Zoom's AI Companion to summarize meetings for attendees) and academic research project that translates natural language commands to programming language using LLMs~\cite{yang2023reactgenie}.
In the above examples, LLMs play a role with different levels of explicitness, which may affect users' mental models and privacy concerns.
In LLM-based systems, a current privacy threat is due to the status quo of API-based development.
Except for big companies like Google that can host their own models, most of these systems incorporate LLMs via web APIs (e.g., OpenAI APIs, open-source LLM endpoints). This means that data sharing with a third party --- the company that hosts the LLM APIs --- may not be clear to users.
In fact, \citet{zhang2023s} discussed an intriguing example in which a user perceived GitHub Copilot as safer than ChatGPT due to the misconception that GitHub Copilot operates entirely on the device.

The interaction modality is another dimension that could potentially impact user privacy.
Prior research has suggested that the human-like interactions provided by LLMs may prompt users to disclose more sensitive information~\cite{zhang2023s, kim2012anthropomorphism, ischen2020privacy}.
As the multimodal LLMs support voice-based or even video-based interactions (e.g., Google Gemini), there is also an open question about how the more streamlined interactions affect users' disclosure behaviors.

\section{Building Usable Tools for Privacy Management for LLM-based Systems}
NLP and system security researchers have investigated privacy-preserving techniques that can be applied throughout the model training and inference phases~\cite{Peris_Dupuy_Majmudar_Parikh_Smaili_Zemel_Gupta_202}.
These backstage strategies may not be helpful for users in managing their context-specific privacy requirements.
This points us to a much needed research direction: developing user-facing privacy management tools for LLM-based systems.
We further categorize it into two problems: 1) helping users sanitize their input; and 2) helping users censor LLM outputs that may contain personal data.

The challenge lies in managing the tradeoff among utility, convenience, and privacy, necessitating interdisciplinary collaborations among HCI, NLP, and system security.
Research with LLM-based CA users has shown that users occasionally remove sensitive information from their input, while this manual process is tedious and easily forgotten~\cite{zhang2023s}.
Sometimes, the task they want to achieve is inherently privacy-sensitive (e.g., personal counseling).
In the absence of support to manage the trade-off between privacy and utility, most of the time, they had to sacrifice privacy completely for utility.
Another privacy/utility trade-off example is the choice between local models and server-based models.

There has been a lot of interest in building personalized LLM agents in HAI and NLP research. However, infusing personal information into LLMs raises significant privacy concerns, especially when the LLMs are used for social tasks that involve other people (e.g., email writing, meeting note taking~\cite{mireshghallah2023can}).
In addition to improving the models' capacities to adhere to privacy norms, there is also a need to build HCI systems that allow users to specify privacy preferences, exercise control, avoid failure cases, and establish trust and accurate mental models of the systems' capabilities.

\section{Challenges and Solutions Beyond Individual Users}

\paragraph{Challenges} 
We have discussed challenges in protecting individual users' privacy.
However, when discussing the human-centered privacy research agenda, we must also consider the backdrop that LLMs have posed challenges to other issues of broad societal importance, such as safety, ethics, and responsible AI use. Solutions to these challenges can conflict with privacy issues.
For example, some levels of monitoring of LLM usage by API providers (e.g., OpenAI) or organizations may be necessary to identify and curtail abusive uses of LLMs --- e.g., create fake news stories, facilitate cybersecurity attacks --- or prevent accidental leakage of proprietary data.
As another example, as ChatGPT and similar language model-based conversational agents make powerful AI more accessible to everyone, they also distribute the obligation to use AI responsibly.
For example, publishers like ACM and Elsevier require authors to transparently report their use of generative AI in their manuscripts.
However, \citet{zhang2023s} highlighted a new privacy concern related to the fear of individuals being found out for using ChatGPT.
This privacy concern could potentially hinder the responsible disclosure of LLM and other generative AI use.
What other considerations need to be taken into account besides safeguarding personal privacy, and how can HCI research assist in understanding and addressing these issues?

\paragraph{Solutions}
HCI and usable privacy researchers have a long history of providing recommendations for policy making and setting industry standards based on solid user research.
To address privacy challenges in everyday applications powered by LLMs, we want to initiate a discussion on how we can maintain our successful track record in engaging with regulatory efforts.
Our aim is to promote the design and development of LLM systems that respect privacy.

Notably, the prevalence of LLMs as a programming assistant tool may also have an impact on the developers' privacy practices.
Research has uncovered security vulnerabilities in LLM-generated code~\cite{pearce2022asleep, asare2023github}, which suggests there may also be an issue with privacy.
Prior research has shown that developers often lack awareness of privacy issues in their code~\cite{li2018coconut}.
Therefore, if privacy issues exist in code generated by LLM, these issues may persist, potentially leading to more widespread privacy violations in general software.

\section{Conclusion}
The advent of large language models (LLMs) has brought about significant privacy challenges.
To build usable, efficient, and privacy-friendly systems powered by these models with imperfect privacy properties, we emphasize the importance of human-centered privacy research in LLMs to complement the extensive model-centered research.
This Special Interest Group (SIG) aims to provide a much-needed space for researchers who are interested in tackling these challenges to openly discuss the problems, research opportunities, research methods, and strategies to collaborate with researchers outside of HCI, such as NLP, system security, and public policy.

\bibliographystyle{ACM-Reference-Format}
\bibliography{sample-base}


\end{document}